\documentstyle[aps,prl,preprint]{revtex}
\begin{document}
\draft
\preprint{Submitted to PRL}
\title{Numerical Renormalization Group Study of Pseudo--fermion
and Slave--boson Spectral Functions in the Single Impurity Anderson Model}
\author
{T. A. Costi, P. Schmitteckert\cite{ps-present}, J.
Kroha\cite{jk-present} and P. W\"{o}lfle}
\address
{Universit\"{a}t Karlsruhe, Institut f\"{u}r Theorie der Kondensierten
Materie, Physikhochhaus, Postfach 6980, 76128 Karlsruhe, Germany}
\maketitle
\begin{abstract}
We use the numerical renormalization group to calculate the auxiliary
spectral functions of the $U=\infty$ Anderson impurity model.
The slave--boson and pseudo--fermion
spectral functions diverge at the threshold with exponents $\alpha_{b}$ and
$\alpha_{f}$ given by the X--ray photoemission and the X--ray
absorption exponents respectively.
In contrast to the NCA, the exact exponents obtained here depend on the
impurity occupation number. In general, vertex
corrections in the convolution formulae for physical Green's functions
are singular at the threshold and may not be neglected in the
Fermi liquid regime.
\end{abstract}
\newpage
Strongly correlated fermi systems have been under intense
investigation over the past few years. This development was triggered
by the discovery of high temperature superconductors, the
subsequent proposals of an electronic mechanism for superconductivity
\cite{anderson.87} and the many fascinating ideas concerning the
normal state \cite{fascinating-ideas}. Despite much theoretical
effort, a systematic and controlled method for dealing with such
systems is still lacking. At the heart of the
problem is the influence of a strong on--site Coulomb repulsion $U$ on a
system of electrons living in a lattice in $d=2$ dimensions. In the
limit of $U\rightarrow\infty$ one has to deal with a projection of the
dynamics into the part of the Hilbert space without any doubly occupied
lattice sites. It is difficult to implement this projection in
conventional perturbation theory. Therefore it has been proposed some
time ago to effect the projection by introducing auxiliary operators
in a larger Fock space \cite{barnes.76}. Thus for spin $\frac{1}{2}$
fermions in a
single band model, where there are three states per lattice site in
the $U\rightarrow\infty$ limit, one introduces a Bose operator
$b_{i}^{\dagger}$ and a pair of Fermi operators
$f_{i\sigma}^{\dagger}$ ($\sigma={\uparrow,\downarrow}$) acting on
their vacuum to create the empty lattice site $i$ and the singly occupied
lattice site $i$ with $\uparrow$ or $\downarrow$. As long as the
number of auxiliary particles is exactly one at each site,
$Q=b_{i}^{\dagger}b_{i}+\sum_{\sigma}f_{i\sigma}^{\dagger}f_{i\sigma}=1$,
the above definition constitutes a faithful representation of the
original model. Unfortunately, the model of interacting bosons and
fermions defined in this way for any physical model of this type (e.g.
the t--J model) is difficult to solve, and few reliable results have
been obtained so far.

In order to develop approximation schemes it is
therefore of interest to apply the auxiliary particle representation
to simpler models, for which comparison with exact results is
possible. One such model is the infinite $U$ single impurity Anderson
model of a magnetic moment in a metal \cite{hewson.93}.

The physics of this model,
in particular the Kondo effect, has been
clarified long ago using qualitative scaling arguments
\cite{scaling.ref}, a numerically
exact renormalization group (NRG) analysis
\cite{nrg.ref}, and the exact solution
based on the Bethe ansatz method \cite{bethe-ansatz.ref}.
A quantitative analytic theory of
the dynamical properties is not completely available yet, although the
numerical renormalization group has been shown to yield close to exact
results for both zero and finite temperature dynamics
\cite{costi.94}. Despite this it is desirable to have an
alternative way of reliably calculating dynamic properties of this
model which is more easily generalized to lattice models. The
slave--boson approach offers such a possibility. It has been
studied within two different approximations. The first one is a
mean--field theory of the bosons and the constraint
field \cite{read.85}. It appears to qualitatively describe the Fermi
liquid behaviour at low temperatures known to exist in this model.
However, it can not describe the smooth cross--over to the free
magnetic moment behaviour at high temperatures. The second one
involves a self--consistent summation of perturbation theory and
exact projection on the physical subspace known
as the ``non--crossing approximation'' or NCA
\cite{nca.ref}. It is found to describe
correctly the cross--over from the high--temperature to the
low--temperature phase, except for the appearance of spurious
non--analytic behaviour in the physical Green's functions below a
characteristic temperature $T_{NCA}$,
where $T_{NCA}<< T_{K}$, the Kondo temperature \cite{mh.84,bickers.87}.
Within the NCA the auxiliary spectral functions for pseudo--fermions
($f$) and slave--bosons ($b$), $A_{f,b}^{+}(\omega)$, above
the threshold, $E_{0}$, behave as $A_{f,b}^{+}(\omega)\sim
|\omega-E_{0}|^{-\alpha_{f,b}}$, with $\alpha_{f}=1/(N+1)$ and
$\alpha_{b}=N/(N+1)$, where $N$ denotes the spin degeneracy. We
note in passing that the corresponding NCA exponents for the
$M$--channel Anderson model have recently been shown to give the correct
singular behaviour \cite{cox.93} expected for $M > 1$.

In order to correctly describe the low--frequency behaviour in such
models, a clear understanding of the infrared
behaviour of the auxiliary spectral functions is essential.
In this paper we address the question of the true behaviour of the
auxiliary spectral functions at zero temperature and in the limit of
frequency $\omega\rightarrow E_{0}$ and clarify the origin of the
non--analyticities in the physical Green's functions within the NCA.
We adapt Wilson's renormalization
group method to the auxiliary particle problem and attempt to
calculate the  spectral functions numerically.

The Anderson model in auxiliary particle representation takes the form
\begin{equation}
H = H_{c} + \epsilon_{d}\sum_{\sigma}f_{\sigma}^{\dagger}f_{\sigma} +
V\sum_{\sigma}(c_{0\sigma}^{\dagger}b^{\dagger}f_{\sigma} + h.c.),\label{eq:AM}
\end{equation}
where
$H_{c}=\sum_{k\sigma}\epsilon_{k}c_{k\sigma}^{\dagger}c_{k\sigma}$ is
the conduction electron kinetic energy and
$c_{0\sigma}=\sum_{k}c_{k\sigma}$ annihilates a conduction electron
with spin $\sigma$ at the impurity site $0$. Following Wilson
\cite{nrg.ref} we (i) linearize the spectrum about the Fermi energy
$\epsilon_{k}\rightarrow k$,
 (ii) introduce a logarithmic mesh of k points $k_{n}=\Lambda^{-n}$
and (iii) perform a unitary transformation of the $c_{k\sigma}$ such
that $c_{0\sigma}$ is the first operator in the new basis and $H_{c}$
takes the form of a tight--binding Hamiltonian in k--space,
\begin{equation}
H_{c} =
\sum_{n=0}^{\infty}\sum_{\sigma}\xi_{n}\Lambda^{-n/2}
(c_{n+1\sigma}^{\dagger}c_{n\sigma}+ h.c.),\label{eq:tight-binding}
\end{equation}
with $\xi_{n}\rightarrow (1+\Lambda^{-1})/2$ for $n>>1$.
These steps are explained in detail in
\cite{nrg.ref}
and can be taken over for the present model without change.

The Hamiltonian (\ref{eq:AM}) together with the discretized form of the
kinetic energy (\ref{eq:tight-binding}) in the new basis is now
diagonalized by the following iterative process: (i) one defines
a sequence of finite size Hamiltonians $H_{N}$ by replacing
$H_{c}$ in (\ref{eq:tight-binding}) by
$H_{N}^{c} =
\sum_{n=0}^{N-1}\sum_{\sigma}\xi_{n}\Lambda^{-n/2}
(c_{n+1\sigma}^{\dagger}c_{n\sigma}+ h.c.)\label{eq:tight-binding-truncated};$
(ii) starting from
$H_{0}=\epsilon_{d}\sum_{\sigma}f_{\sigma}^{\dagger}f_{\sigma} +
V\sum_{\sigma}(c_{0\sigma}^{\dagger}b^{\dagger}f_{\sigma} + h.c.)$,
each successive hopping may be considered
as a perturbation on the previous Hamiltonian; (iii) the Hamiltonians
$H_{N}$ are scaled such that the energy spacing remains the same.
This defines a renormalization group transformation
$\bar{H}_{N+1} = \Lambda^{1/2}\bar{H}_{N} + \xi_{N}
\sum_{\sigma}(c_{N+1\sigma}^{\dagger}c_{N\sigma}+h.c.) - E_{G, N+1}$,
with $E_{G,N+1}$ chosen so that the ground state energy
of $\bar{H}_{N+1}$ is zero. The diagonalization of the Hamiltonians
$\bar{H}_{N}$ is greatly simplified
by the existence of the following conserved quantities: electron
number $N_{e}$, total spin $S$, spin projection $S_{z}$ and auxiliary
particle number $Q$. The eigenstates of $\bar{H}_{N}$ may therefore be
labeled as $|N_{e},S,S_{z},Q;\kappa>$ with $\kappa$ denoting the
remaining quantum numbers, and the Hamiltonian matrices have block
diagonal structure. Considering that the smallest energy scale probed
by $\bar{H}_{N}$ is $\omega_{N} \sim \Lambda^{-(N-1)/2}$, for
a typical Kondo temperature of $T_{K}$ which is $<< 10^{-2}$ of
the conduction band width, it is necessary to iterate
beyond $N\sim 15$ for $\Lambda\approx 2$ to get below the Kondo scale. Since
the dimension of $\bar{H}_{N}$ is growing as $4^{N}$ it is
necessary to truncate the higher energy states for $N>7$. Wilson presents
arguments why this is a valid approximation. We have kept the lowest
250 states in each iteration step and checked that our results remain
unchanged when this number is increased to 2000 states.

In Coleman's slave--boson formulation of the mixed valence problem
\cite{nca.ref} the retarded Green's functions for
pseudo--fermions and slave--bosons are defined in the
enlarged Hilbert space consisting of the disjoint subspaces
$Q=0,1,\ldots$ by
\begin{eqnarray}
G_{f,b}(\omega,T,\lambda) & = & Z_{GC}^{-1}\sum_{Q,m,n}|M_{m,n}^{f,b}|^{2}
(e^{-\beta(\epsilon_{m} + \lambda (Q+1))}
+ e^{-\beta(\epsilon_{n}+\lambda Q)})
/(\omega-(\epsilon_{m}-\epsilon_{n})),\nonumber
\end{eqnarray}
where $-\lambda$ is a chemical potential associated with the
auxiliary particle number $Q$, $Z_{GC}(T,\lambda)$ is the
grand--canonical partition function and
$M_{m,n}^{O}=<Q+1,m|O^{\dagger}|Q,n>$ with $O=f_{\sigma},b$ are the
many--body matrix elements for pseudo--fermions and slave--bosons
respectively. We are interested in the zero temperature projected
spectral functions $A_{f,b}^{+}=-\lim_{T\rightarrow
0}\lim_{\lambda\rightarrow\infty}[\mbox{Im}\;
{G_{f,b}(\omega,T,\lambda)}/\pi]$ and $A_{f,b}^{-}=-\lim_{T\rightarrow
0}\lim_{\lambda\rightarrow\infty}e^{-\beta\omega}[\mbox{Im}\;
{G_{f,b}(\omega,T,\lambda)}/\pi]$,
\begin{eqnarray}
A_{f,b}^{+}(\omega) & = &
\pi\sum_{m}|<1,m|O^{\dagger}|\Phi_{0}>|^{2}
\delta(\omega+E_{GS}^{Q=0}-\epsilon_{m}),\label{eq:aplus}\\
A_{f,b}^{-}(\omega) & = &
\pi\sum_{m}|<\Phi_{1}|O^{\dagger}|0,m>|^{2}
\delta(\omega+\epsilon_{m}-E_{GS}^{Q=1}).\label{eq:aminus}
\end{eqnarray}
Here $|\Phi_{0}>$ is the groundstate of the
$Q=0$ subspace
of non--interacting conduction electrons and $|1,m>$ are the excited states
of the $Q=1$ subspace of the interacting system, $E_{GS}^{Q=0}$
and $\epsilon_{m}$ are the corresponding energy eigenvalues. The spectral
functions $A_{f,b}^{+}$ vanish identically {\em below} the
threshold $E_{0}=E_{GS}^{Q=1}-E_{GS}^{Q=0}$.
Similarly, in the expression for $A_{f,b}^{-}$,
$|\Phi_{1}>$ is the groundstate of the interacting system
($Q=1$ subspace) and $|0,m>$ are the excited states
of the non--interacting conduction electron system, $E_{GS}^{Q=1}$ and
$\epsilon_{m}$ the corresponding energy eigenvalues. These
spectral functions vanish {\em above} the threshold energy $E_{0}$.

The pseudo-particle operators $f_{\sigma}^{\dagger}$, $b^{\dagger}$
in (\ref{eq:aplus}--\ref{eq:aminus}) have matrix
elements $M_{m,n}^{f,b}$ only between
states from the subspace $Q=1$ and the subspace $Q=0$. These matrix
elements are calculated for each iteration step, and are substituted
together with the energy eigenvalue $\epsilon_{m}$ into
(\ref{eq:aplus}--\ref{eq:aminus}) to give spectral functions
$\bar{A}_{N,f,b}^{\pm}(\omega)$.
In principle if all states up to stage $N$ were retained, $H_{N}$
would describe excitations on all energy
scales from the band edge $D=1$ down to the lowest energy scale
present in $H_{N}$, i.e. $\omega_{N}$. Due to the elimination of
higher energy states at each step, the actual range of excitations in
$H_{N}$ is restricted to $\omega_{N}\le \omega \le K\omega_{N}$, where
$K \approx 7$ for $\Lambda \approx 2$ retaining $500$ states
per iteration. Thus at step $N$, the spectral functions are calculated at an
excitation energy $\omega \approx 2\omega_{N}$ in the above range.
The delta functions in (\ref{eq:aplus}--\ref{eq:aminus}) are
broadened with Gaussians of width $\alpha_{N} \approx \omega_{N}$
appropriate to the energy level structure of $H_{N}$.

%
%

The RG calculations were performed for $\Lambda=2$ and
keeping $250$ states per iteration for each subspace ($Q=0,1$).
The hybridization strength $\Delta=\pi\;V^{2}/2D$ was chosen to be
$0.01D$ with the half--bandwidth $D=1$. Several values of the local
level position $\epsilon_{d}$ where chosen in order to characterize the
low energy behaviour of the spectral densities in the various regimes.
In this letter we are mainly concerned with the low frequency limit,
discussion of high energy effects is left to a future publication.

The auxiliary spectral functions exhibit threshold behaviour and
are shown in Fig.\ (\ref{sd-kondo}) in the Kondo
regime with similar behaviour near the threshold in the mixed valent
and empty orbital regimes.
We note that this is not unexpected since
the projected spectral functions (\ref{eq:aplus}--\ref{eq:aminus})
are very similar in appearance to the core--level spectral functions in
the usual X--ray problem so one expects as a result of the
orthogonality catastrophe theorem \cite{anderson.67}
a similar threshold behaviour. This
analogy is useful but requires care since the matrix elements
in (\ref{eq:aplus}--\ref{eq:aminus}) are no longer between
two non--interacting systems as in the X--ray problem.
This leads in particular to a new energy scale, $T_{0}$, for the onset of
the asymptotic power law behaviour, which is $T_{K}$, $\Delta$ or
$\epsilon_{d}$ in the Kondo, mixed valent and empty orbital regimes
respectively. We find that it is only in the Fermi liquid regime,
$|\omega-E_{0}| << T_{0}$, that the power law behaviour is well
characterized.

The results for the threshold exponents are shown in
Fig.\ (\ref{exponents})
as a function of $n_{f}$, the local level occupancy at $T=0$.
The latter was calculated by evaluating $n_{f}(T)$ from the partition
function at a sequence of decreasing temperatures $T_{N}\sim
\Lambda^{-(N-1)/2}$ and then taking the limit $T\rightarrow 0$.
Remarkably these exponents turn out to be the usual photoemission and
absorption exponents for the X--ray problem and are given in terms
of the conduction electron phase--shift at the Fermi level,
$\delta_{\sigma}=\delta_{\sigma}(\epsilon_{F})$, by
\begin{eqnarray}
\alpha_{f} & = & 2\delta_{\sigma}/\pi -
\sum_{\sigma}(\delta_{\sigma}/\pi)^{2} = n_{f} -
n_{f}^{2}/2\label{eq:fermion-exp}\\
\alpha_{b} & = & 1 -
\sum_{\sigma}(\delta_{\sigma}/\pi)^{2} =  1 -
n_{f}^{2}/2\label{eq:boson-exp}
\end{eqnarray}
where the last equations on the RHS of
(\ref{eq:fermion-exp}--\ref{eq:boson-exp}) follow from the
Friedel sum rule, $\delta_{\sigma}=\pi\;n_{f}/2$.
These results are clearly illustrated in Fig.\ (\ref{exponents}) where the
functions $n_{f} - n_{f}^{2}/2$ and $1 - n_{f}^{2}/2$ are plotted
against $n_{f}$ together with the exponents $\alpha_{f,b}$ deduced
from the spectral functions. The exponents $\alpha_{f,b}$ agree with
the RHS of (\ref{eq:fermion-exp}--\ref{eq:boson-exp}) to 3 significant
figures and are the same below and above the threshold,
\begin{equation}
A_{f,b}^{\pm} = a_{f,b}^{\pm}\;|\omega - E_{0}|^{-\alpha_{f,b}}
\label{eq:asymptotic}
\end{equation}
The slave--boson exponent $\alpha_{b}$
corresponds to the usual XPS exponent in the X--ray problem.
On the other hand the pseudo--fermion exponent $\alpha_{f}$ takes the form
of an absorption exponent in the X--ray problem. This is at first
sight unexpected since we are dealing with single--particle Green's
functions in which the processes correspond to removing a particle
from the system altogether. We attempt to give a qualitative
explanation of this. Assuming, as is shown rigorously by the numerical
results, that the threshold exponents are determined by the phase
shifts we note that the removal of a slave--boson has no effect on the
local electronic charge so the relevant phase shifts entering the XPS
exponent are $\delta_{\sigma}$ which are fixed by the
Friedel sum rule. Hence the usual
exponent,$1-\sum_{\sigma}(\delta_{\sigma}/
\pi)^{2}$ , appears in $A_{b}^{\pm}$.
On the other hand the removal of a pseudo--fermion reduces the
electronic charge by one. By charge neutrality the sum of phase shifts
$\sum_{\sigma}\delta_{\sigma}$ is reduced by $\pi$.
Assuming, as in the usual X--ray problem \cite{schotte.69},
that the total phase shift occurs in the channel $\sigma$ in which a
fermion is removed we have that
$1-\sum_{\sigma}(\delta_{\sigma}/\pi)^{2} \rightarrow
1-[(\delta_{\sigma}/\pi - 1)^{2} + (\delta_{-\sigma}/\pi)^{2}]
\rightarrow 2\delta_{\sigma}/\pi -
\sum_{\sigma}(\delta_{\sigma}/\pi)^{2}$
which leads to the exponent $\alpha_{f}$ found numerically.
We note that the same functional form of the exponents on the phase
shift (\ref{eq:fermion-exp}--\ref{eq:boson-exp}) is also found in the
spinless model with constraint. The single phase shift in this case is
given by $\delta=\pi n_{f}$.

With the above results we can gain some insight into the role played by
vertex corrections in the slave--boson method. These appear in Green's
functions for slave--bosons, pseudo--fermions and in the convolution
formulae for physical Green's functions. In general these are
difficult to calculate within perturbative schemes but they may be
estimated within the present non--perturbative approach. First we note
that the above threshold exponents generalized to the N--fold
degenerate model are $\alpha_{f}=2n_{f}/N - n_{f}^{2}/N$ and
$\alpha_{b}=1-n_{f}^{2}/N$ so the NCA exponents $1/(N+1)=1/N +
O(1/N^{2})$ and $N/(N+1)=1-1/N +O(1/N^{2})$ are only correct in the
limit $n_{f}\rightarrow 1$ and $N\rightarrow\infty$. Away from this limit,
vertex corrections in the auxiliary Green's functions, neglected
in the NCA, are therefore important in determining the correct
threshold exponents. To see the influence of vertex corrections on the
physical Green's functions we consider specifically the impurity
Green's function
$G_{\sigma}(\omega)=<<b^{\dagger}f_{\sigma};f_{\sigma}^{\dagger}b>>_{Q=1}$.
This is given in terms of the projected auxiliary
Green's functions, $G_{f,b}$, by
\begin{equation}
G_{\sigma}(i\omega) = -\frac{1}{\beta}\sum_{\nu}G_{f}(i\omega_{\nu} +
i\omega)G_{b}(i\omega_{\nu})V(i\omega_{\nu}+i\omega,i\omega_{\nu})
\label{eq:realgf+vertex}
\end{equation}
where
$V(i\omega_{\nu}+i\omega,i\omega_{\nu})$
represents the renormalized vertex. The effect of vertex
corrections can be seen by evaluating the impurity
spectral density, $\rho_{f\sigma}'(\omega) =
-\mbox{Im}\;[G_{\sigma}(\omega+i\delta)/\pi]$, at low frequencies
without vertex corrections, i.e. for $V=1$. Using
(\ref{eq:fermion-exp}--\ref{eq:asymptotic}) in
(\ref{eq:realgf+vertex}) gives
\begin{equation}
\rho_{f\sigma}'(\omega\rightarrow 0^{+}) =
a^{+}_{f}a^{-}_{b}\omega^{1-\alpha_{f}-\alpha_{b}}
B(1-\alpha_{f},1+\alpha_{b})\label{eq:singularity}
\end{equation}
where $B$ is the Beta function. Thus in the absence of vertex
corrections the impurity spectral density diverges at the Fermi level
as a power law
$|\omega|^{1-\alpha_{f}-\alpha_{b}}=|\omega|^{-n_{f}(1-n_{f})}$.
 From the Friedel sum rule,
$\rho_{f\sigma}(\omega=0)=\sin^{2}(\pi n_{f}/2)$, so
the vertex corrections are singular at low energies, i.e. close to the
threshold,
and lead to a singularity in $\rho_{f\sigma}$ at the Fermi level
which cancels that in (\ref{eq:singularity}).
This singular behaviour together with next leading order corrections to the
spectral functions and vertex corrections should restore the Fermi liquid
behaviour which is well known for this model. Similar vertex
corrections appear in the calculation of other physical quantities
such as the dynamic spin susceptibility.

%
%

To summarize, we have characterized the low energy behaviour of the auxiliary
spectral functions of the single--channel Anderson model at $T=0$.
The power law behaviour close to the threshold is approached slowly and only
sets in well below the characteristic low energy scale of the model.
The threshold exponents were related to the conduction electron phase
shifts and depend on the occupation number $n_{f}$, in contrast to the
corresponding NCA exponents. Vertex corrections in the pseudo--fermion
and slave--boson Green's functions are therefore required
to obtain the correct threshold exponents. They are
also needed in the convolution formulae for physical Green's functions
to restore the well known Fermi liquid behaviour for the
single--channel model.
It is interesting to speculate whether the vertex corrections are
non--singular at the threshold in multi--channel
impurity models exhibiting non--Fermi liquid behaviour.
Work on this is in progress and will be reported elsewhere.
The results presented here could provide guidance in developing
reliable self--consistent slave--boson schemes. These could then be extended
to the important case of lattice models, which are more difficult to
study within the numerical renormalization group approach.

\acknowledgments
We are grateful to T. Kopp for useful discussions.
This work was supported by E.U. grant no. ERBCHRX CT93 0115 (TAC, PW),
the Alexander von Humboldt Foundation (JK) and the Deutsche
Forschungsgemeinschaft (PW, PS).


\begin{figure}
\caption{
The pseudo--fermion $A_{f}^{\pm}$ (solid lines) and slave--boson
$A_{b}^{\pm}$ (dashed lines)
spectral functions in the Kondo case $\epsilon_{d}=-6\Delta$,
$T_{K}/\Delta = 8.07\times 10^{-4}$
(using $T_{K}=\Delta^{1/2}e^{-\pi\epsilon_{d}/2\Delta}$),
$n_{f}=0.934$. The $+$ signs
are for the spectral function above the threshold, $E_{0}$,
and the circles are for the spectral function below the threshold.
The arrow indicates the position of $|\epsilon_{d}|$.}
\label{sd-kondo}
\end{figure}
\begin{figure}
\caption{
The exponents $\alpha_{f,b}$ deduced from the asymptotic power
law behaviour of the auxiliary spectral functions as calculated within
the NRG for different values of the occupation $n_{f}$. The solid
lines are the functions $n_{f}-n_{f}^{2}/2$ ($\diamond$) and
$1-n_{f}^{2}/2$ ($\circ$).
}
\label{exponents}
\end{figure}
\end{document}